\documentclass[runningheads,a4paper]{llncs}

\usepackage{amssymb}
\usepackage{graphicx}
\usepackage{algorithmic}
\usepackage{algorithm}

\usepackage{tikz}
\usetikzlibrary{calc}
\usetikzlibrary{patterns}
\tikzstyle{mybox} = [draw=black, fill=white,  thick,
    rectangle, inner sep=10pt, inner ysep=20pt]
\tikzstyle{mybox} = [draw=black, fill=white,  thick,
    rectangle, inner sep=2pt, inner ysep=2pt]

\usepackage{setspace}

\begin{document}


\title{On Planar Visibility Counting Problem}
\author{Sharareh Alipour}
\institute{School of computer science, Institute for research in fundamental sciences(IPM)}
\maketitle

\begin{abstract}
For a set $S$ of $n$ disjoint line segments in $\mathbb{R}^{2}$, the visibility counting problem is to preprocess $S$ such that the number of visible segments in $S$ from any query point $p$ can be computed quickly. There have been approximation algorithms for this problem with trade off between space and query time. We propose a new randomized algorithm to compute the exact answer of the problem. For any $0<\alpha<1$, the space, preprocessing time and query time are $O_{\epsilon}(n^{4-4\alpha})$, $O_{\epsilon}(n^{4-2\alpha})$ and $O_{\epsilon}(n^{2\alpha})$, respectively. Where $O_{\epsilon}(f(n)) = O(f(n)n^{\epsilon})$ and $\epsilon>0$ is an arbitrary constant number.
\end{abstract}
\section{Introduction}

Let $S=\{s_1, s_2,..., s_n\}$ be a set of $n$ disjoint closed line segments in the plane contained in a bounding box. Two points $p,q$ in the bounding box are visible to each other with respect to $S$, if the line segment $\overline{pq}$ does not intersect any segment of $S$. A segment $s_i\in S$ is also said to be visible with respect to $S$ from a point $p$, if there exists a point $q\in s_i$ such that $q$ is visible to $p$. \textit{The Visibility Counting Problem (VCP)} is to find $m_p$, the number of segments of $S$ visible from a query point $p$. We know that \textit{the visibility region of a given point $p\in \mathbb{R}^{2}$} is defined as \textit{VP}$_{S}(p) = \{ q\in \mathbb{R}^{2}: p$ and $q$ are visible with respect to $S$ $\}$,  and \textit{the visibility region of a given segment $s_i$} is defined as \textit{VP}$_{S}(s_i) = $\{$p\in \mathbb{R}^{2}: s_i$ and $p$ are visible with respect to $S$ $\}$. Note that the size of \textit{VP}$_{S}(p)$ is $O(n)$, but there are examples that the size of \textit{VP}$_{S}(s_i)$ is $\Theta(n^4)$.

Consider the $2n$ end-points of the segments of $S$ as vertices of a geometric graph. Add a straight-line-edge between each pair of visible vertices. The result is \textit{the visibility graph of $S$} or \textit{VG(S)}. We can extend each edge of \textit{VG(S)} in both directions to the points that the edge hits some segments in $S$ (or the bounding box). This generates at most two new vertices and two new edges. Adding all these vertices and edges to \textit{VG(S)} results in a new geometric graph, called \textit{the extended visibility graph of $S$} or \textit{EVG(S)}. We can use \textit{EVG(S)} to compute the visibility region of any segment $s_i\in S$~\cite{gud}.

\subsection{Related Works}

There is an $O(n\log n)$ time algorithm that can compute \textit{VP}$_S(p)$ using $O(n)$  space~\cite{asa,sur}. Vegter propose an output sensitive algorithm that reports \textit{VP}$_S(p)$ in $O\left(|\textit{VP}_S(p)|\log\left(\frac{n}{|\textit{VP}_S(p)|}\right)\right)$ time, by preprocessing the segments in $O(m\log n)$ time using $O(m)$ space, where $m=O(n^{2})$ is the number of edges of \textit{VG(S)} and $|\textit{VP}_S(p)|$ is the number of vertices of $\textit{VP}_S(p)$~\cite{vet}. We can also solve VCP using \textit{EVG(S)}. Consider the planar arrangement of the edges of \textit{EVG(S)} as a planar graph. All points in any face of this arrangement can see the same number of visible segments and this number can be computed for each face in the preprocessing step~\cite{gud}. Since there are $O(n^{4})$ faces in the planar arrangement of \textit{EVG(S)}, a point location structure of size $O(n^{4})$ can answer each query in $O(\log n)$ time. However, $O(n^4)$ preprocessing time and space is high. We also know that by computing $\textit{VP}_S(p)$, any query can be answered in $O(n\log n)$ with no preprocessing. This has led to several results with a tradeof between the preprocessing cost and the query time~\cite{aro,bos,poc,zar} (please refer to~\cite{gho2} for a complete survey). 

Suri and O'Rourke~\cite{sur} introduce the first 3-approximation algorithm for VCP. Their algorithm is based on representing a region by the union of a set of convex (triangular) regions. Gudmundsson and Morin~\cite{gud} improve this result to a 2-approximation algorithm using an improved covering scheme. For any $0<\alpha\leq1$ their method builds a data structure of size $O_{\epsilon}(m^{1+\alpha})=O_{\epsilon}(n^{2(1+\alpha)})$ using $O_{\epsilon}(m^{1+\alpha})=O_{\epsilon}(n^{2(1+\alpha)})$ preprocessing time, from which each query is answered in $O_{\epsilon}(m^{(1-\alpha)/2})=O_{\epsilon}(n^{1-\alpha})$ time, where $O_{\epsilon}(f(n)) = O(f(n)n^{\epsilon})$ and $\epsilon>0$ is an arbitrary constant number. This algorithm returns $m'_{p}$ such that $m_{p}\leq m'_{p}\leq 2m_{p}$. The same result can be achieved by using the algorithms described in~\cite{ali}~and~\cite{nor}. In \cite{ali}, considering the endpoints of the segments, it is proven that the number of visible end-points, denoted by $ve_p$ is a 2-approximation of $m_p$, which means $m_p\leq ve_p\leq 2m_p$. Another notable approximation algorithm for VCP is described by Fischer \textit{et al.}~\cite{fis1,fis2} to estimate the saving of applying a visibility algorithm for computer graphics applications. 

Alipour~\textit{et al.}~\cite{alil} propose two randomized approximation algorithms for VCP. The first algorithm depends on two constants $0\leq \beta\leq \frac{2}{3}$ and $0<\delta\le 1$, and the expected preprocessing time, the expected space, and the expected query time are $O(m^{2-3\beta/2}\log m)$, $O(m^{2-3\beta/2})$,  and $O(\frac{1}{\delta^2}m^{\beta/2}\log m)$, respectively. In the preprocessing phase, the algorithm selects a sequence of random samples, whose size and number depend on the computation time and memory space tradeoff parameters. When a query point $p$ is given by an adversary unaware of the random sample of our algorithm, it computes the exact number of visible segments from $p$, denoted by $m_p$, if $m_p\leq \frac{3}{\delta^2}m^{\beta/2}\log(2m)$. Otherwise, it computes an approximate  value, $m'_p$, such that with the probability of at least $1-\frac{1}{m}$, we have $(1-\delta)m_p\leq m'_p\leq (2+2\delta)m_p$.
The preprocessing time and space of the second algorithm are $O(n^2\log n)$ and $O(n^2)$, respectively. This algorithm computes the exact value of $m_p$ if $m_p\leq \frac{1}{\delta^2}\sqrt{n}\log n$. Otherwise, it returns an approximate value $m''_p$ in expected $O(\frac{1}{\delta^2}\sqrt{n}\log n)$ time, such that with the probability at least $1-\frac{1}{\log n}$, we have $(1-3\delta)m_p\leq m''_p\leq (1.5+3\delta)m_p$.

Alipour~\textit{et al.}~\cite{alip} study the problem over terrain. Given a $2.5D$ terrain and a query point $p$ on or above it, they propose an approximation algorithm to find the triangles of terrain that are visible from $p$. They implement and test their algorithm on real data sets.

\subsection{Our Result}

We combine the algorithms described in~\cite{ali} and~\cite{gud} to propose a randomized algorithm to find the exact answer of VCP. The expected space and the preprocessing time of our algorithm are $O_{\epsilon}(m^{2-2\alpha})=O_{\epsilon}(n^{4-4\alpha})$ and $O_{\epsilon}(m^{2-\alpha})=O_{\epsilon}(n^{4-2\alpha})$, respectively and the expected query time is $O_{\epsilon}(m^{\alpha})=O_{\epsilon}(n^{2\alpha})$. Our algorithm returns the exact value of  VCP, as opposed the previous approximation algorithms.
Where $0< \alpha <1$ and $m=O(n^2)$ is the number of edges in the \textit{EVG(S)}.
\section{Preliminaries}

In this section, we present  theorems and results that we utilize in our algorithm.

\begin{theorem} 
\textnormal{(\cite{alin})}
\label{vt}
For a set of $n$ disjoint line segments, we can preprocess them in $O_{\epsilon}(n^2)$ time using $O_{\epsilon}(n^2)$ space such that in expected $O(n^{\epsilon})$ query time, we can check whether a given query point $p$ and a given segment $s_i\in S$ can see each other or not.  
\end{theorem}

According to \cite{gud}, it is possible to cover the visibility region of each segment $s_i\in S$ with $O(m_{s_i})$ triangles denoted by $\textit{VT}(s_i)$ such that $|\textit{VT}(s_i)|=O(m_{s_i})$, where  $m_{s_i}$ is the number of edges of \textit{EVG(S)} incident on $s_i$. Note that the visibility triangles of $s_i$ may overlap. If we consider the visibility triangles of all segments, then there is a set $\textit{VT}_S=\{\Delta_1,\Delta_2,...\}$ of $|\textit{VT}_S|=O(m)$ triangles, where $m=O(n^2)$ is the number of edges in \textit{EVG(S)} \cite{gud}. We say $\Delta_i$ is related to $s_j$ if and only if $\Delta_i\in \textit{VT}(s_j)$.

Gudmundsson and Morin proved that for a given query point $p$, the number of triangles in $\textit{VT}_S$ containing $p$ is between $m_p$ and $2m_p$~\cite{gud}. So, for any query point~$p$, they compute the number of triangles containing $p$ to give a 2-approximation solution of VCP. 

Let $T=\{\Delta_1,\Delta_2, \ldots, \Delta_n\}$ be a set of $n$ triangles. For any $n'$ with $n\leq n'\leq n^2$, the \textit{multi-level partition tree} data structure of size $O_{\epsilon}(n')$  can be constructed in $O_{\epsilon}(n')$ time. By using the multi-level partition tree, the number of triangles containing a query point $p$ can be counted in $O_{\epsilon}(n/\sqrt n')$ time and the triangles can be reported by $O(k)$ extra query time where $k$ is the number of reported triangles (see Theorem~1 in \cite{gud}). Please refer to \cite{aga,mat} for the details of the partition tree data structure and its efficient implementation. 

In the algorithm proposed by Gudmundsson and Morin~\cite{gud}, there are $O(m)$ visibility triangles. By $O_{\epsilon}(m^{1+\alpha})=O_{\epsilon}(n^{2(1+\alpha)})$ preprocessing time and space, each query is answered in $O_{\epsilon}(m^{(1-\alpha)/2})=O_{\epsilon}(n^{1-\alpha})$ time, where $0<\alpha\leq1$. Obviously, if we color the visibility triangles of each segment $s_i$ with color $c_i$, then the number of triangles with distinct colors containing $p$ is the number of segments visible from $p$. Note that we can use their algorithm to find the exact number of triangles with distinct color containing $p$, but the query time depends on the value of $m_p$, which is not efficient for the query points with a large set of visible segments.

\section{The Algorithm}

Before we describe our algorithm, we explain a concept usually used in algorithms with a divide-and-conquer approach. Let $L$ be a set of $n$ lines in the plane and $r\leq n$ be a given parameter. A $(1/r)$-cutting for $L$ is a collection of (possibly unbounded) triangles with disjoint interiors, which covers all the plane and the interior of each triangle intersects at most $n/r$ lines of~$L$. The first (though not optimal) algorithm for constructing a cutting is given by Clarkson~\cite{cla}. He proved that a random sample of size $r$ of the lines of $L$ can be used to produce an $O(\log r/r)$-cutting of size $O(r^2)$. The efficient construction of $(1/r)$-cuttings of optimal size $O(r^2)$ can be found in \cite{deb1} and \cite{cha}. The set of triangles in the cutting together with the collection of lines intersecting each triangle can be found in $O(nr)$ time.

We use the algorithm described in \cite{cla} for generating an $(1/r)$-cutting. Let $E=\{e_1,e_2,...,e_{m'}\}$, $m'=O(m)$, be the set of edges of all the visibility triangles. So, we can partition the space into $r^2$ triangular regions such each region is intersected by at most $m'\log r/r$ edges of $E$. Let $r=m^{1-\alpha}$. Let $R=\{R_1,R_2,...,R_{m''}\}$, $m''=O(m^{{2-2\alpha}})$, be the set of triangular regions in our cutting. So each triangular region is crossed by at most $m^{\alpha} \log r$ edges of $E$.

Now we describe our algorithm based on the $(1/r)$-cutting approach. In the preprocessing stage, we choose a random point $p_i$ from each triangular region $R_i$. We then compute the number of triangles in $T$ with distinct colors containing $p_i$ and denote it by $mp_i$. This is actually is the same as the number of segments that are visible from $p_i$. Note that $T$ is the set of given $O(m)$ colored triangles. For each $R_i$, we record $m_{p_i}$ and the location of $p_i$. Thus, the plane is partitioned into $m''$ regions and we record a point $p_i$ and the number of triangles with distinct colors in $T$ containing $p_i$ for each region. This data structure needs $O(m^{{2-2\alpha}})$ space. We have the following theorem for computing the segments that a given query segment intersects.

\begin{theorem}
\emph{\cite{deb}}
\label{cs}
Let $S$ be a set of $n$ segments in the plane and $n\leq n'\leq n^2$, we can preprocess the segments in $O_{\epsilon}(n')$ time such that for a given query segment $s$, the segments crossed by $s$ are reported in $O_{\epsilon}(n/\sqrt n' +k)$, where $k$ is the number of segments crossed by $s$.
\end{theorem}

In the preprocessing stage, we construct the data structure described in Theorem~\ref{cs} on the edges of $T$. The set  $T$ has $O(m)$ edges. Let $n'=m^{2-2\alpha}$. So, the preprocessing time and space for this data structure is $O_{\epsilon}(m^{2-2\alpha})$ and for a given segment $s$, we can report the segments of $T$ crossing $s$ in $O_{\epsilon}(m^{\alpha})$ time.

In the query time, for a given query point $p$, first we find the region $R_i$ containing $p$ in $O(\log m)$ time. Then, we find the edges of triangles that cross $\overline{pp_i}$. The number of these edges crossing $R_i$ is $O(m^{\alpha})$, so the number of edges that cross $\overline{pp_i}$ is $O(m^{\alpha})$.
Thus,  we can report these edges in $O_{\epsilon}(m^{\alpha})$ using $O_{\epsilon}(m^{2-2\alpha})$ preprocessing time and space (see~Theorem~\ref{cs}).

\begin{figure}[htpb]
	\centering
	\begin{tikzpicture}[scale=0.6]
\draw(0,0)--(1,3)[red];
\draw(1,3)--(3,1)[red];
\draw(3,1)--(0,0)[red];

\draw(-2,4)--(0,-1)[red];
\draw(0,-1)--(3,0)[red];
\draw(3,0)--(-2,4)[red];

\draw(-2,0)--(2,1)[red];
\draw(2,1)--(2,-1)[red];
\draw(2,-1)--(-2,0)[red];

\draw(3,3)--(1,-3)[green];
\draw(1,-3)--(-4,1)[green];
\draw(-4,1)--(3,3)[green];

\draw(1,0)--(5,4)[blue];
\draw(5,4)--(6,-1)[blue];
\draw(6,-1)--(1,0)[blue];

\draw(-1.2,-0.2)--(-1.3,4)[blue];
\draw(-1.3,4)--(5,1)[blue];
\draw(5,1)--(-1.2,-0.2)[blue];

\draw(0,7)--(7,0)[dashed];

\draw(0,7)--(-7,-6)[dashed];
\draw(-7,-6)--(7,7)[dashed];

\draw(7,0)--(-7,-6)[dashed];
\draw(7,0)--(4,-7)[dashed];
\draw(4,-7)--(3.27,3.78)[dashed];
\draw(4,-7)--(-7,-6)[dashed];

\filldraw(-1,0) circle(2pt);
\draw (-1,0) node[above] {$p_i$};
\filldraw(-1,3) circle(2pt);
\draw (-1,3) node[above] {$p$};

\filldraw(-2.5,0) circle(2pt);
\draw (-2.5,0) node[above] {$R_i$};

\end{tikzpicture}

\caption{A sample visibility scenario. Here, $p_i$ is selected for the region $R_i$. The number of triangles with distinct color containing $p_i$ is 3. In the preprocessing stage, this value is computed for each $p_i$. For the query point $p$, we consider the set of segments that intersect $\overline{pp_i}$. The initial answer for $p$ is $m_{p_{i}}$. Because both $p_i$ and $p$ are inside a red triangle, we do not change the value of $m_{p_{i}}$ for red triangles. Because $p_i$ is inside a green triangle and $p$ is not inside the same green triangle, the value of $m_{p_i}$ is reduced by 1 and the final answer is~$2$. 
}
\label{color}
\end{figure}
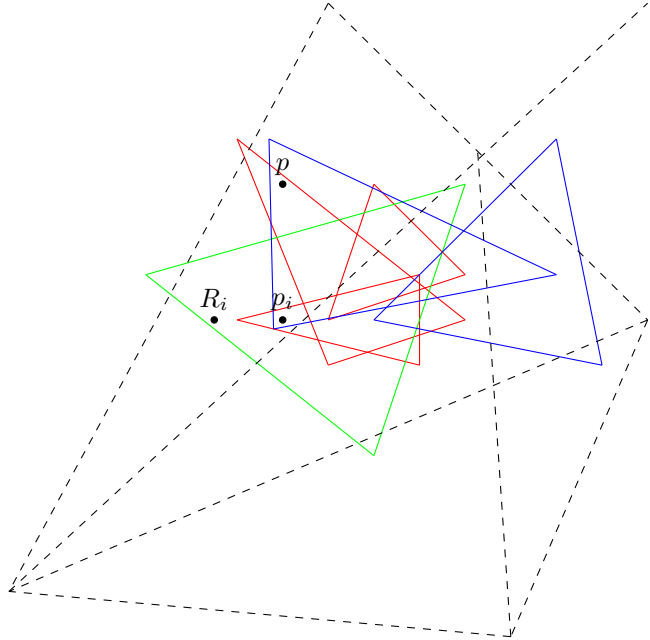

For each $p_i$, we know the number of triangles with distinct colors containing $p_i$ that is the number of segments visible from $p_i$. We also know that the difference between the number of visible segments from $p$ and $p_i$ can be computed by considering the edges of $T$ that cross $\overline{pp_i}$. So, to compute the number of visible segments from $p$, we compute the distinct colors of edges that cross $\overline{pp_i}$. The expected number of these colors is $O(m^{\alpha})$, which means they are related to $O(m^{\alpha})$ segments of $S$. For any segment $s_i$ that at least one of the edges of the visibility triangles of $s_i$ crosses $\overline{pp_i}$, we use Theorem~\ref{vt} to test whether $p_i$ and $p$ are visible from $s_i$. If $s_i$ is not visible from $p_i$ and is visible from $p$, then we add $1$ to the value of $m_{p_{i}}$, where $m_{p_{i}}$ is the number of visible segments from $p_i$. If $s_i$ is visible from $p_i$ and is not visible from $p$, then we subtract $1$ from $m_{p_{i}}$. At the end, we return the new value of $m_{p_i}$ that is the number of segments visible from $p$. Note that we test at most $O(m^{\alpha})$ segments. For each segment $s_j$, we use Theorem~\ref{vt} for testing the visibility of ($s_j$, $p$) and ($s_j$, $p_i$). So, the query time is $O_{\epsilon}(m^{\alpha})$. See Figure~\ref{color} for a sample visibility scenario. So, we conclude the following theorem.

\begin{theorem}
\label{l1}
The VCP can be answered in expected $O_{\epsilon}(m^{\alpha})$ computation time with the expected $O_{\epsilon}(m^{2-2\alpha})$ memory space.
\end{theorem}

\subsection*{Preprocessing time}
The preprocessing time of this algorithm depends on computing $m_{p_{i}}$. We start with $R_1$. First, we compute $m_{p_{1}}$ in $O(n\log n)$ and then move to one of the neighbor regions of $R_1$, for example $R_2$. We can compute $m_{p_{2}}$ in $O(m^{\alpha})$ time. Since there are $O(m^{2-2\alpha})$ regions, the preprocessing times is $O_{\epsilon}(m^{\alpha}.m^{2-2\alpha})=O_{\epsilon}(m^{2-\alpha})$.

\section{Conclusion}

We proposed an exact algorithm for the visibility counting problem. By exploiting the results described in~\cite{gud}, we transform the problem into computing the triangles with distinct colors containing a query point, which is known as general range search problem. The algorithms for general range search problem are output sensitive~\cite{aga1,pro}. However, the nature of the visibility counting problem enabled us to solve the problem in a query time independent of the size of output. In our general range search problem for a given query point $p$ the difference between $p$ and $p_i$ depends on the edges of the triangles that cross $pp_i$. Each edge $e$ was related to a visibility triangle of a segment $s$. So, the processing of $e$ was possible by testing the visibility of $(p,s)$ and $(p_i,s)$  in expected time of $O(\log n)$.

As a future work suppose that we are given some triangles in $R^3$ and the goal is to compute the number of visible triangles from a query point $p$.
Our approach can be extended to solve the problem in $R^3$.

Gudmundsson and Morin~\cite{gud} proposed a $2$-approx\-imation factor algorithm for $VCP$. We produce the exact answer of VCP with the same storage and query time of their algorithm. However, our preprocessing time is higher than that of their algorithm. So, an interesting open question is to answer the problem with the same preprocessing time of their algorithm.



\small
\bibliographystyle{abbrv}

\bibliography{visibility}

\end{document}